# Practical Quantum Cryptography: The Q-KeyMaker™.

Fabio Antonio Bovino, Maurizio Giardina

*Quantum Optics Lab, Selex Sistemi Integrati, via Puccini 2, 16154, Genova*

*Abstract* — **In the next years the data transmission connections will constitute one of the principal tools of communication among cities, enterprises and public administration. With their enhanced connectivity, the systems and nets of information are now exposed to an increased vulnerability and new safety problems are emerging. Insofar Quantum Key Distribution (QKD) has matured to real world applications and can enhance the safety of the communication nets. In this paper we present the QKD network designed and implemented by Selex-SI and we give an overview of the obtained results.**

*Keyword:* **Quantum Cryptography**

## I. Introduction

IN the next years the data transmission connections will constitute one of the principal tools of communication among cities, enterprises and public administration. Personal computers more and more powerful, convergent technologies and an ample use of Internet have replaced the precedents "autonomous systems" with limited working abilities, predominantly restricted to closed nets. Today, the interested parties are widely interconnected and the connections overcome the national confinements. Besides, Internet is the support for vital infrastructures as energy, transports and financial activities, and it revolutionized the way in which enterprises manage their own activities, the governments assure the services to the citizens and the enterprises and the citizens communicate and exchange information. The nature of the technologies that constitute the infrastructure of communications and information have recorded a similar notable evolution. So the typologies, the volume and the sensitive character of the exchanged information are increased in a substantial way. With their enhanced connectivity, the systems and nets of information are now exposed to an increased number and a wider range of threats and vulnerabilities and, therefore, new safety problems are emerging. Insofar the cryptography and the safety of the nets are matured, bringing to the development of practical and easy to use applications. Nowadays, the Protection of the Net can use both standard algorithms (DES, 3-DES, RSA) and proprietary algorithms. Keys of encryption of suitable length are used to guarantee the strength of both the Reservation (Strong Encryption) and the Authentication (Strong Authentication). Nevertheless, the oldest and more delicate problem connected to the conventional systems of encryption remains the **management of the keys**. In fact, this is an essential component of a good operation of every cryptographic system, and it includes the generation, the possible maintenance and the dispatch to the legitimate recipients. In each of these phases a security loss can jeopardize the whole structure. In other words the management of the keys represents the true Achilles' heel of every military or civil cryptographic system. The keys owe therefore to undergo two antithetical and contradictory demands: **to stay secret and protected,** and **to be replaced as more frequently as be possible**.

A key of encryption consists of a string of casual bits. Such string is useful only if the perfect randomness is assured, it is known only to the authorized users and it is regularly renewed.

Quantum Cryptography permits to solve all these problems, allowing the distribution of couples of identical keys in a sure way, guaranteeing their perfect randomness and, therefore, providing the One-Time-Pad encryption [1-3]. This new technology can play a fundamental role for the protection of the so-called Critical National Infrastructures (CNI), that are a key component of the national security in numerous countries, and, in the United States, an issue at the centre of the debate on terrorism and Security since September 11.

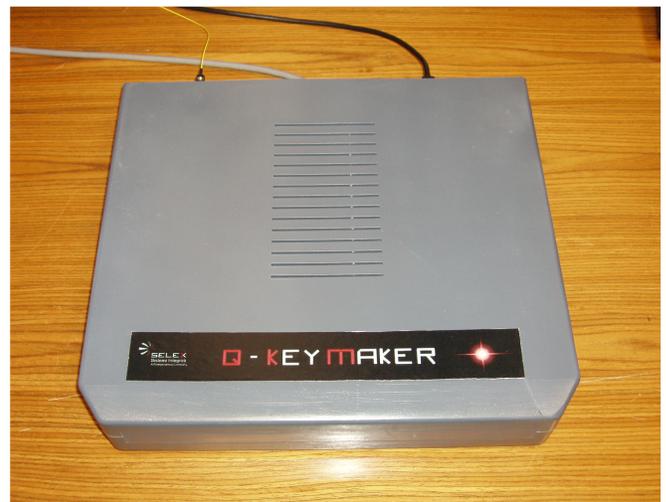

Figure 1. The Q-KeyMaker.

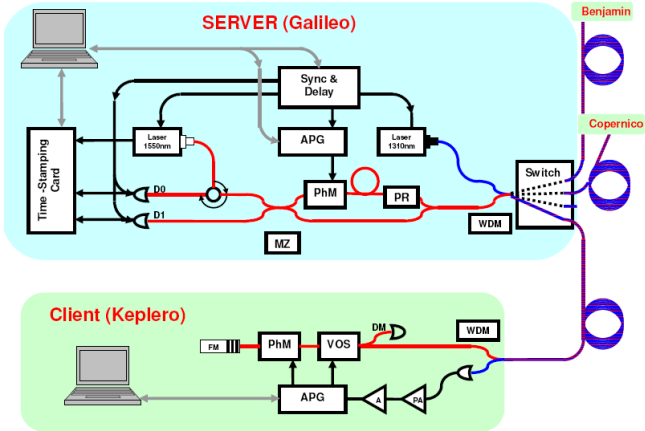

Figure 2. The QKD network (see text for details).

CNI are in fact based on Critical Information Infrastructures (CII). CII must allow the correct CNI working under normal conditions of operation, guarantee a suitable operational ability in case of critical events, and assure an elevated degree of security. Moreover, the recent developments in the quantum technologies, particularly of Quantum Computers, constitute a sword of Damocle on the modern cryptography and one possible use of them is destined to bring a great revolution.

Here we present the Q-KeyMaker, a four-user QKD network in a star configuration, i.e. with one server (Galileo) and three clients (Benjamin, Copernico and Keplero), based on decoy-state method.

## II. PRACTICAL QUANTUM CRYPTOGRAPHY WITH DECOY STATE: THE Q-KEYMAKER

Practical QKD systems can differ in many important aspects from their original theoretical proposal that typically demand technologies that are beyond our present experimental capability. Especially, the signals emitted by the source, instead of being single photons, are usually Weak Coherent Pulses (WCP), with typical average photon numbers of 0.1 or higher. The quantum channel introduces errors and considerable attenuation (about 0.2dB/km) that affect the signals even when Eve is not present. Besides, for telecom wavelengths, standard InGaAs single-photon detectors can have detection efficiency below 15% and are noisy due to dark counts. All these differences reduce the security of the protocols, and lead to limitations of the rate and the maximum distance that can be covered by these techniques [4]. A main security threat of practical QKD schemes based on WCP arises from the fact that some signals contain more than one photon prepared in the same state. For this reason Eve is no longer limited by the no-cloning theorem [5] since in these events the signal itself provides her with perfect copies of the signal photon. She can perform, for instance, the so called photon number splitting (PNS) attack on the multi-photon pulses [4]. This attack gives Eve full information about the part of the key generated with the multi-photon signals, without causing any disturbance in the signal. As a result, it turns out that the standard BB84 protocol [6] with WCP can deliver a key generation rate of order $O(\eta^2)$, where $\eta$ ($<<1$) denotes the transmission efficiency of the quantum channel [7, 8]. To achieve higher secure key rates over longer distances, different QKD schemes, that are robust against the PNS attack, have been proposed in recent years [9 -13]. One of these schemes is the so-called decoy state QKD [9 -11] where Alice varies, independently and at random, the mean photon number of each signal state that she sends to Bob, by employing different intensity settings.

Eve does not know a priori the mean photon number of each signal state sent by Alice. This means that her eavesdropping strategy can only depend on the photon number of these signals, but not on the particular intensity setting used to generate them. From the measurements corresponding to different intensity settings, the legitimate users can estimate the classical joint probability distribution describing their outcomes for each photon number state. This gives them a better estimation of the behavior of the quantum channel, and enhances the achievable secret key rate and distance. This technique has been successfully implemented in several recent experiments [14], and it can give a key generation rate of order $O(\eta)$ [9 - 11].

The Selex-SI Q-KeyMaker is a plug and play auto-compensating system (see Fig.2). It employs a pulse ($\lambda$=1550 nm) emitted by Galileo's laser diode at a frequency of 4 MHz. The pulse is split at a first 50/50 Beam-Splitter (BS). The two resulting pulses impinge the two input ports of a Polarizing Beam Splitter (PBS), after traveling, respectively, through a short and a long arm of an unbalanced interferometer (MZ). The linear polarization is turned by $90^0$ in the long arm, so that the two pulses exit the PBS through the same port. The separated pulses travel down to one of the three clients, selected by a micro mirror optical switch, they are reflected on a Faraday Mirror and come back orthogonally polarized and codified in phase by a Phase Modulator (PhM). A Polarization Independent LiNbO3 switch (VOS) then allows to prepare and control the Signal state (with an average number of photons equal to μ) and the two Decoy states (ν and vacuum). The optimal value of μ and ν is chosen taking into account the length of the single mode fiber link (15km Benjamin, 21km Copernico and 25Km Keplero) and the current QBER of the channel (see Fig.3-4).

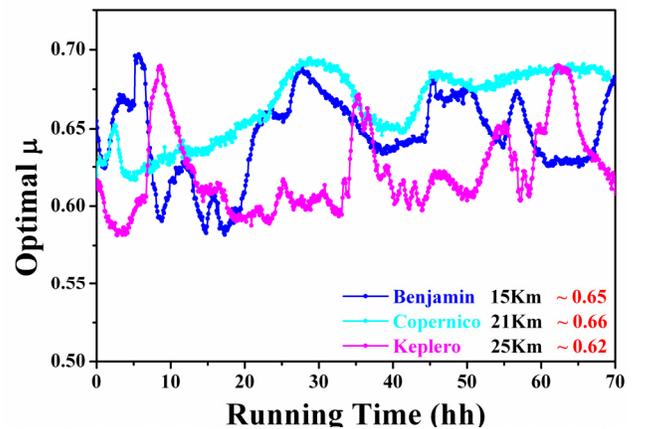

Figure 3. The optimal value µ is chosen taking into account the length of the single mode fiber link and the current QBER of the channel.

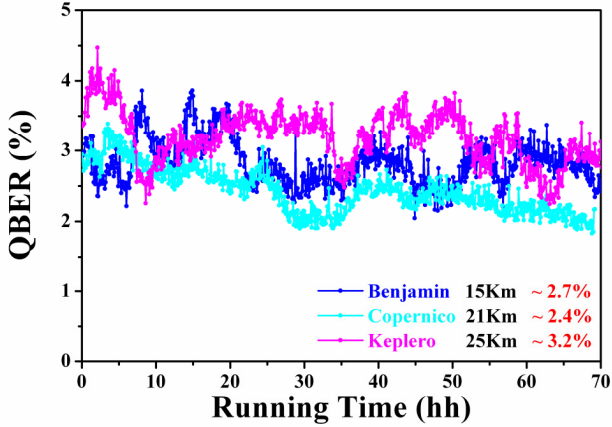

Figure 5. QBER behavior for the different links.

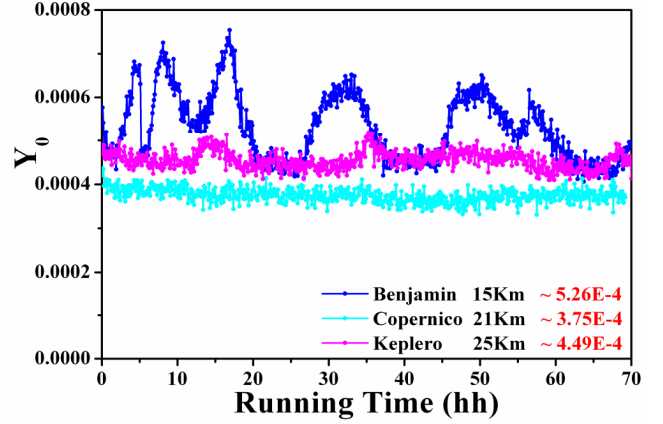

Figure 4. Background noise for the different links.

In turn, both pulses take now the other path at Bob's interferometer and arrive simultaneously at the first BS, where they interfere. Finally they are detected by InGaAs Avalanche Photon Detectors ($D_0$, $D_1$). A clock signal, synchronized with the transmitter, controls the measurement basis choice for the phase modulator located in the unbalanced fiber Mach-Zender (MZ) interferometer on the server's side, the gating signals (2.5ns) for the InGaAs detectors $D_0$ and $D_1$ (15% quantum efficiency) and a second pulsed laser (λ =1310 nm) for client's device synchronization.

Decoy pulse QKD theory gives a rigorous bound for the characteristics of the single photon pulses, which are the only source pulses that contribute to the secure bit rate. In decoy protocol, three different average pulse intensities (referred to as Signal, Decoy pulse and Vacuum) with mean photon numbers μ, ν (μ > ν) and 0 are used. Our setting for the proportion of three transmitted states is 14:1:1 among the signal state, decoy state and vacuum state. By using the results of [13, 15] we obtain the following key generation rate:

$$R \geq q\{Q_\mu f(E_\mu)H_2(E_\mu) + Q_1[1 - H_2(e_1)]\}, \quad (1)$$

$Q_\mu = Y_0 + 1 - e^{-\eta\mu}$ and $E_\mu$ are, respectively, the measured gain and the quantum bit error rate (QBER) for the signal state; q is an efficiency factor for the protocol (q=0.5 for standard BB84); the $H_2(x)$ is the binary entropy function, while the factor $f(x)$ takes into account the efficiency of the bidirectional error correction; $Q_1$ and $e_1$ are the unknown gain and the error rate of the true single photon state in signal states. To achieve the maximum possible key generation rate, the decoy state method gives an estimation of the lower bound of $Q_1$ denoted as $Q^{L_1}$, and the upper bound of $e_1$ denoted as $e^{U_1}$. After experimentally measuring all the relevant parameters, we can input the following bounds for calculating the final key generation rate

$$Q_1 \geq Q^{L_1} = \frac{\mu^2 e^{-\mu}}{\mu\nu - \nu^2}\left[Q_\nu e^\nu - Q_\mu e^\mu \frac{\nu^2}{\mu^2} - \frac{\mu^2 - \nu^2}{\mu^2}Y_0\right], \quad (2)$$

and

$$e_1 \leq e^{U_1} = \frac{E_\nu Q_\nu - \varepsilon_0 Y_0 e^{-\nu}}{Q^{L_1}}, \quad (3)$$

where $Y_0$ (see Fig.5) and $\varepsilon_0$ represent, respectively, the background and the error rate (~0.5) when Vacuum state is set; $E_\nu$ and $Q_\nu = Y_0 + 1 - e^{-\eta\nu}$ are the error and the gain for the Decoy State. The telecom single mode fibers have an average attenuation of about 0.2 dB per kilometer. Besides the transmission loss in the fiber, there are also other coupling and connection losses, principally 1.7 dB due to the polarization maintaining fiber in server's side, optical switch and fiber connectors. The 70hh data set (see Fig.6-7) is a part of 150 days working and monitoring period during which similar results have been obtained: this shows the high reliability and stability of Q-KeyMaker. The Secure Key Generation Rate was of 5.8Kb/s for 25Km link (Keplero, Qber 3.2%), 8 Kb/s for 21Km link (Copernico, Qber 2.4%) and 11.5Kb/s for 15km link (Benjamin, Qber 2.7%).

### III. QKD NETWORK

Few groups, such as BBN, have demonstrated quantum networks with three nodes or more. Phoenix et al. [16] proposed the concept of passive quantum networks. Such a network adopts passive optical couplers as network nodes to split photons sent by the transmitter, to more receivers. In this architecture, simultaneous communication, or "broadcast", from one user to all others is established, and quantum key distribution between one user and any other one can be realized. Townsend et al. [17] conceptually demonstrated that quantum key distribution is feasible between any couple of users within a passive quantum network. However, in the passive communication network, photons (and hence the bit that they represent) are split by couplers according to their ratio so that only a subset of these photons reach each terminal node. Therefore, the actual key rate between specific terminals is greatly reduced. In comparison, in the active QKD Selex-SI network configuration a controller actively drives an optical switch to set the communication direction (orientation). In this case, all photons emitted by Bob (except those lost in the link) are delivered to the selected terminal, yielding the full communication speed (therefore the highest quantum key rate) between two users in the network. The one-to-any structure is a natural extension of a four-node architecture obtained by replacing the 1 x 3 with a 1 x n optical switch. For a any-to-any architecture, optical cross connection (OXC) switches could be used to implement communications between any Alice and any Bob. After each switching operation, a time alignment procedure is necessary, which sums up to the communication time.

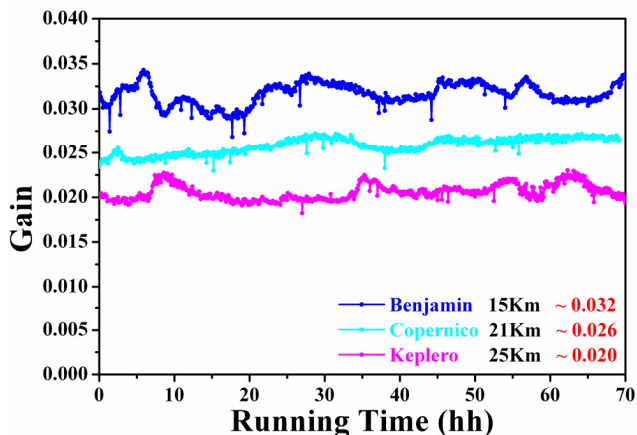

Figure 6. Signal Gain for the different links.

## IV. CONCLUSIONS

In summary, we reported the demonstration of an operational network QKD system: the full key exchange and application protocols were performed in real time among three clients and one server. The generated quantum keys can be immediately employed for ciphering applications.

ACKNOWLEDGMENT

We acknowledge the financial support from the Italian Defense Ministry; Contr. 9223 di Rep. in data 22.12.2005, "Programma di ricerca tecnologica per la Crittografia Quantistica Avanzata".

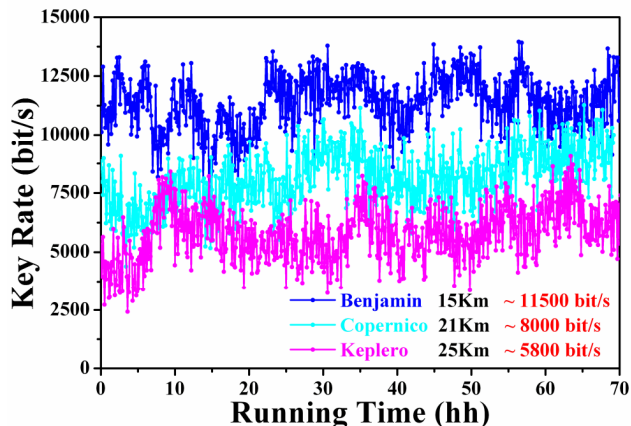

Figure 7. Secure Key Rate for the different links.